\documentclass[twocolumn,prb]{revtex4}
\usepackage{amsfonts}
\usepackage{times}
\usepackage{amsmath,amsbsy,amssymb,graphicx}
\usepackage{color}

\newcommand{\red}[1]{{\textcolor{red}{#1}}}
\let\emph=\red

\begin{document}

\title{Chiral anomaly enhancement and photo-irradiation effects in
multi-band touching fermion systems}
\author{Motohiko Ezawa}
\affiliation{Department of Applied Physics, University of Tokyo, Hongo 7-3-1, 113-8656,
Japan}

\begin{abstract}
Multi-band touchings together with the emergence of fermions exhibiting
linear dispersions have recently been predicted and realized in various
materials. We first investigate the Adler-Bell-Jackiw chiral anomaly in
these multi-band touching semimetals when they are described by the
pseudospin operator in high dimensional representation. By evaluating the
Chern number, we show that the anomalous Hall effect is enhanced depending
on the magnitude of the pseudospin. It is also confirmed by the analysis of
the Landau levels when magnetic field is applied. Namely, charge pumping
occurs from one multi-band touching point to another through multi-channel
Landau levels in the presence of parallel electric and magnetic fields. We
also show a pair annihilation of two multi-band touching points by
photo-irradiation. Furthermore, we propose generalizations of Dirac
semimetals, multiple-Weyl semimetals and loop-nodal semimetals to those
composed of fermions carrying pseudospins in high dimensional
representation. Finally we investigate the 3-band touching protected by the $%
C_{3}$ symmetry. We show that the 3-band touching point is broken into two
Weyl points by photo-irradiation.
\end{abstract}

\maketitle

\section{Introduction}

Weyl semimetal has created one of the most active fields of modern condensed
matter physics\cite{Hosur,Jia}. Experimental observations\cite{Xu,Lv}
accelerate the study. It is described by a two-band theory with a linear
dispersion. It is topologically protected by the monopole charge\cite%
{Murakami}. A prominent feature is the emergence of the Adler-Bell-Jackiew
(ABJ) anomaly\cite{ABJ}, where the chiral charge is not preserved in the
presence of the parallel magnetic and electric fields\cite%
{Zyuzin,Aji,Goswami,Para,Chen,Liu}. This anomalous phenomenon has
experimentally been observed\cite{CZhang,CZhang2,Huang,Jun,Li}. The pair
creation or annihilation of Weyl semimetals by photo-irradiation has also
been discussed\cite{PWang,PChan, Ebihara,PYan, PChan2}.

Recently, there are reports on new types of fermions which have no
counterparts in particle physics. They appear at multi-band touching points
with linear dispersions in crystals, which we may call multi-band touching
fermions. Examples have been found at 3-band\cite%
{Brad,Chang,HWeng,HWeng2,Zhu}, 4-band\cite{Ezawa4}, 6-band\cite{Brad} and
8-band\cite{Brad,Wieder} touching points. Among them, there are proposals
that 3-band\cite{Brad} and 4-band\cite{Ezawa4} touching fermions are
described by the pseudospin operator in $J$-dimensional representation with $%
J=1$ and $3/2$, respectively. Let us refer to this kind of multi-band
touching fermions as $J$-fermions. Their experimental observations are yet
anticipated. On the other hand, there is another kind of 3-band touching
fermions appearing along the C$_{3}$ symmetry invariant line, which is
protected by the C$_{3}$ symmetry\cite{Chang,HWeng,HWeng2,Zhu}. We refer to
them as the C$_{3}$-protected fermions. They have experimentally been
observed\cite{Ding}.

In this paper, motivated by these discoveries, we propose and explore
various models composed of multi-band touching fermions. The minimal models
are obtained by replacing the Pauli matrices\ with the pseudospin operators
having a high dimensional representation. In this minimal scheme we may
easily generalize Weyl semimetals, multiple Weyl semimetal, Dirac
semimetals, and loop-nodal semimetals to those composed of $J$-fermions for
a general value of $J$.  We analyze the Landau levels (LLs) and
photo-irradiation effects to reveal underlying physics. We also study the
photo-irradiation effects of the C$_{3}$-protected fermions.

This paper is composed as follows. In Section II, we investigate a system
made of fermion multiplets described by the pseudospin operator $\mathbf{J}$
appearing at a multi-band touching point. Each band, indexed by the
eigenvalue $j$ of the operator $\mathbf{J}$, is shown to have a monopole
with the charge $2j$ in the momentum space. We introduce the Chern number by
counting the monopole charges associated with the bands below the Fermi
energy. The Chern number contributes to the anomalous Hall conductivity in
the 3-dimensional (3D) space. Then, we analyze the Landau levels (LLs) by
applying external magnetic field to the multi-band touching system.

In Section III, we introduce a Hamiltonian describing a pair of $J$%
-fermions. Here we recall that in a crystal multi-band touching points
always appear in pairs, whose chiralities are opposite\cite{NN}. When
electric field is applied additionally in parallel with the magnetic field,
certain LLs convey electric charges from one multi-band touching point to
another, as is a manifestation of the ABJ anomaly. The ABJ anomaly is
closely related to the Chern number. Next, we study the effect of
photo-irradiation applied to the multi-band touching semimetal. We show that
a pair-annihilation of two $J$-fermions is induced by photo-irradiation.

In Sec IV, V and VI, we generalize Dirac fermions, multiple-Weyl semimetals
and loop-nodal semimetals by replacing the Pauli matrices with the
pseudospin operators. Especially, we find various transitions to occur in
the generalized Dirac-like $J$-fermions by photo-irradiation.

In Section VII, we show that the C$_{3}$-protected fermion is broken by
photo-irradiation. We have prepared Supplementary Material, where various
formulas used in this work are derived in details.

\section{$J$-fermions}

\textbf{Hamiltonian:} The emergence of multi-band touching is assured at
high-symmetry points by crystalline symmetry in some lattice models\cite%
{Brad,Ezawa4}. In the vicinity of a multi-band touching point the effective
Hamiltonian is well described by%
\begin{equation}
H=\hbar v\mathbf{k}\cdot \mathbf{J},  \label{kJ}
\end{equation}%
where examples are known for $J=1$ and $3/2$. Here, $\mathbf{J}%
=(J_{x},J_{y},J_{z})$ is the generator of the pseudospin with magnitude $J$, 
$J\geq 1/2$. Note that $J$-fermions emerge at a touching point of $2J+1$
bands at the zero energy. It can be viewed as the generalization of the Weyl
fermion described by the Hamiltonian $H=\hbar v\mathbf{k}\cdot \mathbf{%
\sigma }$ with the Pauli matrix $\mathbf{\sigma }$. We investigate the
Hamiltonian (\ref{kJ}) for general $\mathbf{J}$.

The Hamiltonian (\ref{kJ}) can be fermionic or bosonic in general. However,
in the bosonic case, the bosons are condensed around the bottom of the band.
Namely, the occupancy in the vicinity of the multi-band touching point is
scarce and the multi-band touching point becomes inactive. On the other
hand, in the fermionic case, the Fermi sea can be set at the multi-band
touching point due to the Fermi degeneracy and the multi-band touching point
plays an important role. In the following, we only consider multi-band
touching fermions.

The Lie algebra reads $\left[ J_{\mu },J_{\nu }\right] =i\varepsilon _{\mu
\nu \rho }J_{\rho }$. Using the polar coordinate, $k_{x}=k\sin \theta \cos
\phi ,\quad k_{y}=k\sin \theta \sin \phi ,\quad k_{z}=k\cos \theta $, the
Hamiltonian is diagonalized as $UHU^{-1}=\hbar vkJ_{z}$ with 
\begin{equation}
U=\exp [i\theta J_{x}]\exp [i\left( \phi +\frac{\pi }{2}\right) J_{z}].
\end{equation}%
Using the relation%
\begin{equation}
U^{-1}J_{z}U=J_{x}\sin \theta \cos \phi +J_{y}\sin \theta \sin \phi
+J_{z}\cos \theta ,
\end{equation}%
the energy spectrum is given by 
\begin{equation}
E=j\hbar vk,
\end{equation}%
where $j$ labels the band with $j=-J,-J+1,\cdots ,J-1,J$. The eigenstates
functions are given by $\left\vert \psi _{j}\right\rangle =U^{-1}\left\vert
j\right\rangle $.

\textbf{Monopole charges:} With the use of the eigenstate $\left\vert \psi
_{j}\right\rangle $, the Berry curvature is rewritten as%
\begin{eqnarray}
-i\mathbf{\Omega }_{j} &=&\frac{\mathbf{k}}{k^{3}\sin \theta }\left[
\left\langle \frac{\partial \psi _{j}}{\partial \theta }\right\vert \left. 
\frac{\partial \psi _{j}}{\partial \phi }\right\rangle -\left\langle \frac{%
\partial \psi _{j}}{\partial \phi }\right\vert \left. \frac{\partial \psi
_{j}}{\partial \theta }\right\rangle \right]  \notag \\
&=&\frac{\mathbf{k}}{k^{3}\sin \theta }\left\langle j\right\vert \frac{%
\partial U}{\partial \theta }\frac{\partial U^{-1}}{\partial \phi }-\frac{%
\partial U}{\partial \phi }\frac{\partial U^{-1}}{\partial \theta }%
\left\vert j\right\rangle .
\end{eqnarray}%
Using the relations%
\begin{equation}
\frac{\partial U}{\partial \theta }\frac{\partial U^{-1}}{\partial \phi }-%
\frac{\partial U}{\partial \phi }\frac{\partial U^{-1}}{\partial \theta }%
=J_{z}\sin \theta -J_{y}\cos \theta ,  \notag
\end{equation}%
the Berry curvature is explicitly calculated for each band as%
\begin{equation}
\mathbf{\Omega }_{j}=i\nabla \times \left. \left\langle \psi _{j}\right\vert
\nabla \psi _{j}\right\rangle =j\frac{\mathbf{k}}{k^{3}}.
\end{equation}%
Since $\frac{1}{2\pi }\int d^{3}k\,\nabla \cdot \mathbf{\Omega }_{j}=2j$,
the Berry curvature of the band indexed by $j$ describes a monopole carrying
the monopole charge $2j$ in the momentum space. This result is already known
for Weyl semimetals\cite{Murakami}, 3-band\cite{Brad} and 4-band\cite{Ezawa4}
touching cases. We have shown that it is a generic result in the system (\ref%
{kJ}).

\textbf{Chern numbers:} We may associate the Chern number to the multi-band
touching points. Since it is defined only in the 2D space, we treat $k_{z}$
as a parameter and analyze the system of 2D fermions on the $k_{x}$-$k_{y}$
space for each $k_{z}$. Namely we evaluate the Chern number for the band $j$
as a function of $k_{z}$ as 
\begin{equation}
C_{j}\left( k_{z}\right) =\frac{1}{2\pi }\iint dk_{x}dk_{y}\Omega _{j}^{z}=-j%
\text{sgn}\left( k_{z}\right) .
\end{equation}%
The $k_{z}$-dependent total Chern number is given by taking the sum under
the Fermi energy, 
\begin{equation}
C\left( k_{z}\right) =\sum_{j\leq 0}C_{j}\left( k_{z}\right) =\frac{1}{2}N%
\text{sgn}\left( k_{z}\right) ,
\end{equation}%
where%
\begin{equation}
N=-2\sum_{j\leq 0}j=\left\{ 
\begin{array}{lll}
\left( J+1/2\right) ^{2} & \text{for} & J\in \text{half integer},\smallskip
\\ 
J\left( J+1\right) & \text{for} & J\in \text{integer}.%
\end{array}%
\right.  \label{FormuN}
\end{equation}%
For examples, $N=1$ for $J=1/2$, $N=2$ for $J=1$, $N=4$ for $J=3/2$ and $N=6$
for $J=2$. The Chern number $C(k_{z})$ is quantized. The sign changes at $%
k_{z}=0$, where a multi-band touching point exists.

\begin{figure}[t]
\centerline{\includegraphics[width=0.5\textwidth]{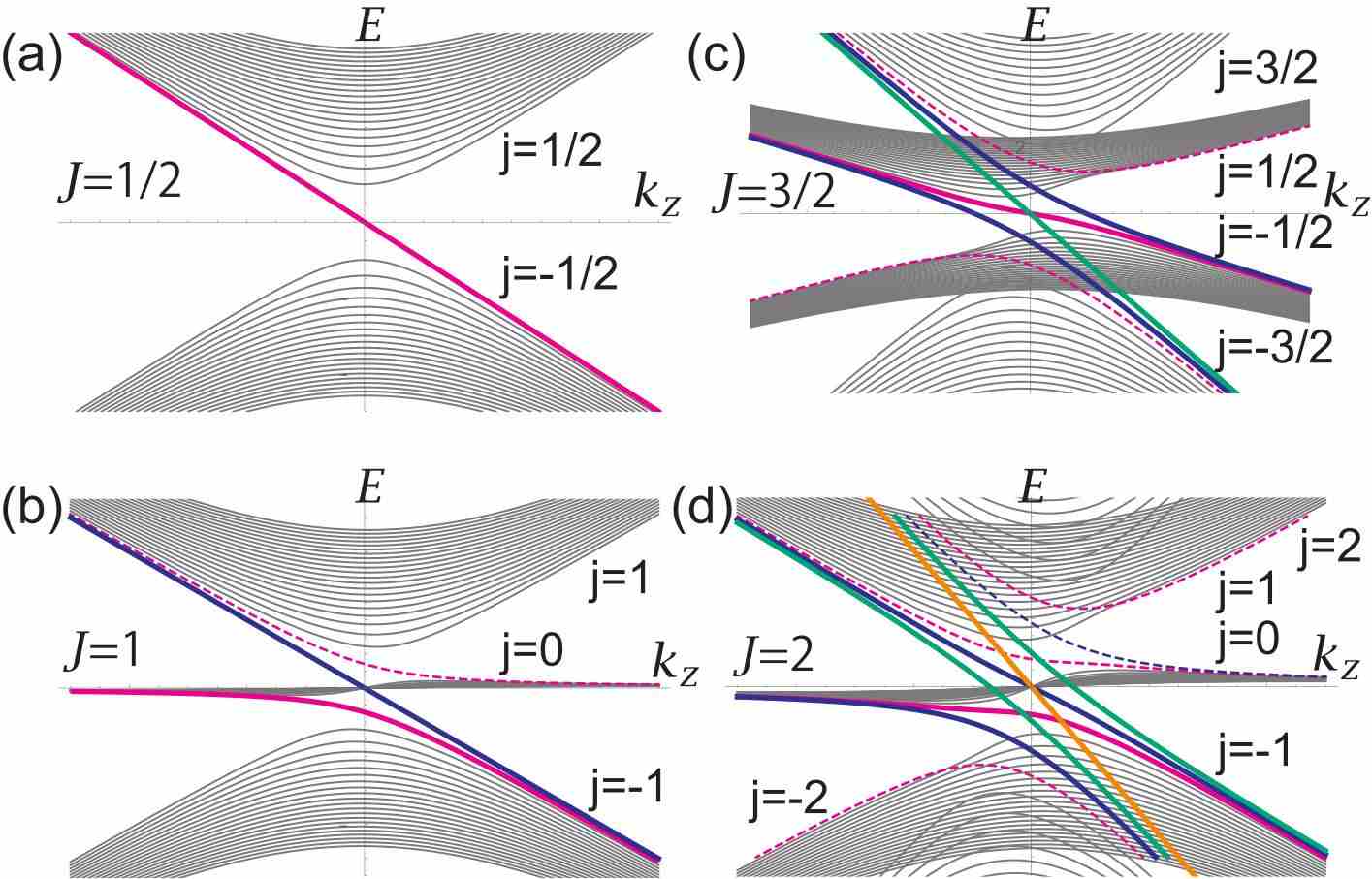}}
\caption{ Landau Levels for a multi-band touching fermion multiplet. The
horizontal axis represents the momentum $k_z$, while the vertical axis
represents the energy $E$. The energy spectrum of the LLs is determined as a
function of $k_{z}$. The spectrum resembles that of a nanoribbon, which
consists of the bulk spectrum and the edge modes (colored curves). We refer
to the LLs corresponding to edge modes as the mid-gap LLs. Mid-gap LLs
connecting the bands with $\Delta j=1$ are marked in magenta, those with $%
\Delta j=2$ are marked in cyan, those with $\Delta j=3$ are marked in green
and those with $\Delta j=4$ are marked in orange. The mid-LLs contributing
to the ABJ anomaly are emphasized in thick curves, while the other mid-LLs
are indicated by dotted curves. (a) Weyl semimetal with $J=1/2$, (b) 3-band
touching semimetal with $J=1$, (c) 4-band touching semimetal with $J=3/2$
and (d) 5-band touching semimetal with $J=2$. }
\label{FigLL}
\end{figure}

\textbf{Landau levels of J=1 fermions:} We proceed to include a homogeneous
magnetic field $\mathbf{B}=\mathbf{\nabla }\times \mathbf{A}=\left(
0,0,-B\right) $ with $B>0$ along the $z$ axis into the multi-band touching
fermion system. Landau-levels for the $J=1$ fermion are discussed in Ref.%
\cite{Brad}. Let us summarize it in order to generalize it to the $J$%
-fermions. By making the minimal substitution to the Hamiltonian (\ref{kJ}),
we obtain%
\begin{equation}
\hat{H}=\hbar \omega _{\text{c}}\left( \hat{a}^{\dagger }J_{-}+\hat{a}%
J_{+}\right) +\hbar vk_{z}J_{z}
\end{equation}%
with the covariant momentum $P_{i}\equiv \hbar k_{i}+eA_{i}$. We have
introduced a pair of Landau-level ladder operators, 
\begin{equation}
\hat{a}=\ell _{B}(P_{x}+iP_{y})/\sqrt{2}\hbar ,\quad \hat{a}^{\dagger }=\ell
_{B}(P_{x}-iP_{y})/\sqrt{2}\hbar ,
\end{equation}%
satisfying $[\hat{a},\hat{a}^{\dag }]=1$, where $\ell _{B}=\sqrt{\hbar /eB}$
is the magnetic length and $\omega _{\text{c}}=\sqrt{2}\hbar v/\ell _{B}$.

There are two types of spectra; the bulk spectrum and the mid-gap spectrum.
See the caption of Fig.\ref{FigLL}. First, we obtain the bulk spectrum by
setting 
\begin{equation}
\psi =\left( u_{n}\left\vert n\right\rangle ,u_{n+1}\left\vert
n+1\right\rangle ,u_{n+2}\left\vert n+2\right\rangle \right) ^{t}
\end{equation}%
for $n\geq 0$ and solving the eigenvalue problem $\hat{H}\psi =E\psi $.
There are three solutions since there are three variables $u_{n}$, $u_{n+1}$
and $u_{n+2}$. They produce three LLs. Second, we obtain the mid-gap
spectrum as follows. There are two solutions by setting $\psi =\left(
0,u_{0}\left\vert 0\right\rangle ,u_{1}\left\vert 1\right\rangle \right)
^{t} $, yielding two LLs, as illustrated by two magenta curves in Fig.\ref%
{FigLL}(b). One connects the bands between $j=1$ and $j=0$ and the other
connects the bands between $j=0$ and $j=-1$. We also have one LL by setting $%
\psi =\left( 0,0,u_{0}\left\vert 0\right\rangle \right) ^{t}$, which
connects the band between $j=1$ and $j=-1$, as illustrated by one cyan curve
in Fig.\ref{FigLL}(b). This LL has zero energy at $k_{z}=0$. These LLs
connects two bulk bands in all combinations, and then the number of the
mid-gap LLs is given by $N_{\text{mid}}={{}_{3}C_{2}}=3$.

\textbf{Landau levels of J=3/2 fermions:} We may carry out a similar study
for the case with $J=3/2$. First, we obtain the bulk spectrum by setting 
\begin{equation}
\psi =\left( u_{n}\left\vert n\right\rangle ,u_{n+1}\left\vert
n+1\right\rangle ,u_{n+2}\left\vert n+2\right\rangle ,u_{n+3}\left\vert
n+3\right\rangle \right) ^{t}
\end{equation}%
for $n\geq 0$. Next, we obtain the mid-gap spectrum: There are three LLs by
solving $\psi =\left( 0,u_{0}\left\vert 0\right\rangle ,u_{1}\left\vert
1\right\rangle ,u_{2}\left\vert 2\right\rangle \right) ^{t}$ connecting the
bands with $\Delta j=1$, two LLs by solving $\psi =\left(
0,0,u_{0}\left\vert 0\right\rangle ,u_{1}\left\vert 1\right\rangle \right)
^{t}$ connecting the bands with $\Delta j=2$ and one LL by solving $\psi
=\left( 0,0,0,u_{0}\left\vert 0\right\rangle \right) ^{t}$ connecting the
bands with $\Delta j=3$. There are $N_{\text{mid}}={{}_{4}C_{2}}$ $=6$ LLs.
We show the LLs as a function of $k_{z}$ in Fig.\ref{FigLL}(c).

On the other hand, the number of the LLs contributing to the ABJ anomaly is $%
N_{\text{ABJ}}=4$ since the two bands connecting ($j=3/2$ and $j=1/2$) and ($%
j=-3/2$ and $j=-1/2$) do not cross the Fermi energy. This is consistent with
the result that the anomalous Hall conductance becomes $N$ times larger than
that of the Weyl semimetal with $N=4$ for $J=3/2$: See (\ref{AnomaHC}).
Namely, $N_{\text{ABJ}}=N$.

\textbf{Landau levels }of $J$-fermions: We generalize the above scheme to
the system of $J$-fermions. The number of the mid-gap LLs is $N_{\text{mid}%
}=\sum_{i=1}^{2J}=2J(2J+1)/2$. It is also derived as $N_{\text{mid}}={%
{}_{2J+1}C_{2}}$ by the fact that the mid-gap LLs connect all the bands.

On the other hand, the LLs contributing to the ABJ anomaly are counted as%
\begin{equation}
N_{\text{ABJ}}=N_{\text{mid}}-2_{J+1/2}C_{2}=\left( J+1/2\right) ^{2}
\end{equation}%
for half integer $J$, and%
\begin{equation}
N_{\text{ABJ}}=N_{\text{mid}}-2_{J+1}C_{2}-J=J\left( J+1\right)
\end{equation}%
for integer $J$, where we have subtracted the LLs which are away from the
Fermi energy. Comparing these with (\ref{FormuN}), we find the relation $N_{%
\text{ABJ}}=N$. See Fig.\ref{FigLL}(d) for the case of $J=2$. Recall that $%
N_{\text{ABJ}}=1$ for the Weyl semimetal with $J=1/2$ [Fig.\ref{FigLL}(a)].
Consequently the ABJ anomaly is $N$ times enhanced compared with the Weyl
semimetal.

\begin{figure}[t]
\centerline{\includegraphics[width=0.5\textwidth]{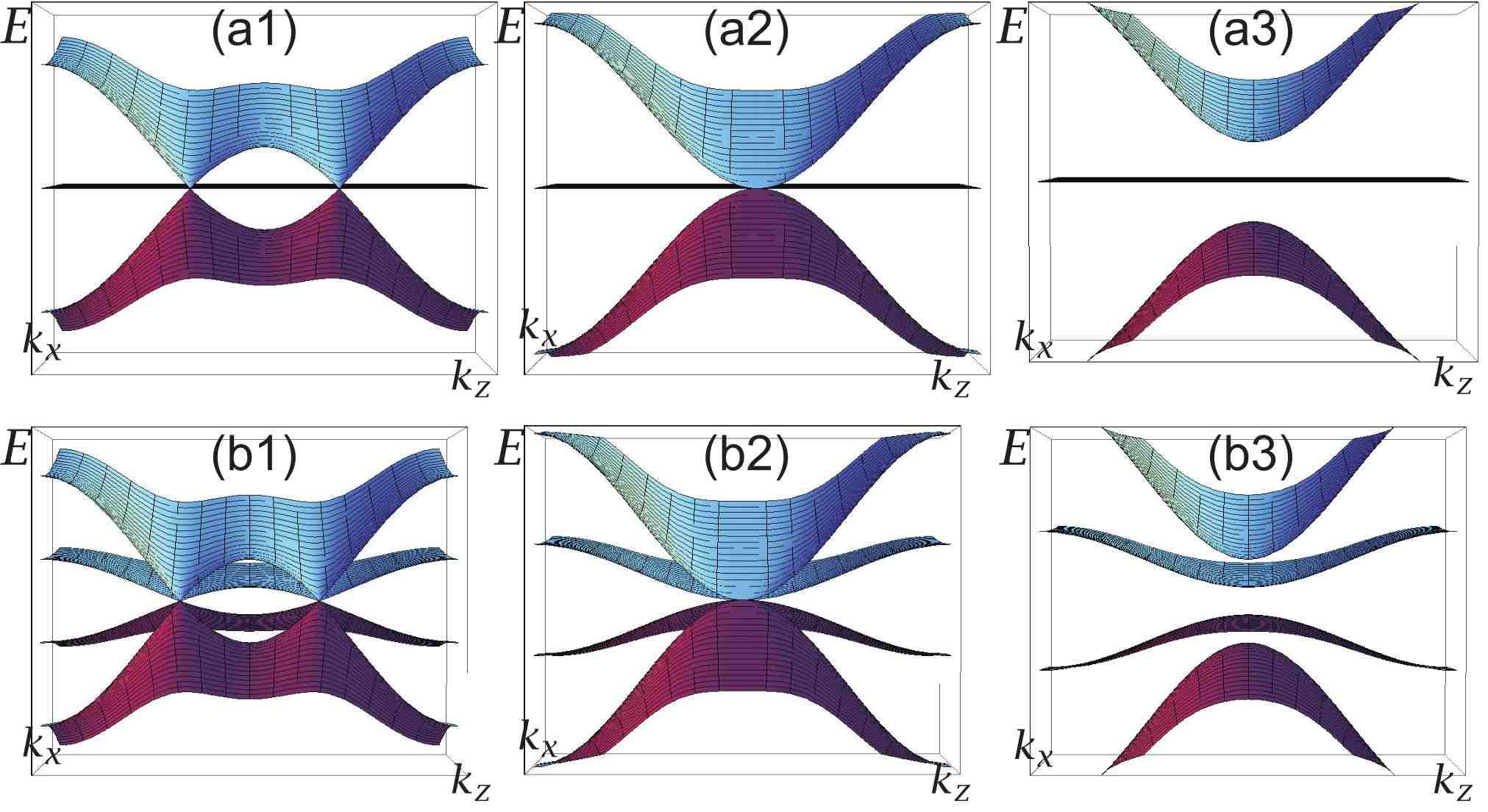}}
\caption{ Pair annihilation of a pair of multi-band touching multiplets. The
horizontal plane is spanned by the $k_{z}$ and $k_{x}$ axes. The vertical
axis is the energy $E$ as a function of $k_{z}$ and $k_{x} $, where we have
set $k_{y}=0$. (a1) 3-band touching semimetal with $J=1$ without
photo-irradiation. (a2) That with the critical photo-irradiation. (a3) That
above the critical point. (b1), (b2), (b3) The corresponding ones for 4-band
touching semimetal with $J=3/2$. }
\label{FigPhoto}
\end{figure}

\section{Pair of $J$-fermions}

In crystals, multi-band touching points always appear in pairs, whose
chiralities are opposite\cite{NN}. A simplest model which has a pair of
multi-band touching points is described by the Hamiltonian 
\begin{equation}
H=\hbar vk_{x}J_{x}+\hbar vk_{y}J_{y}+\frac{\hbar v}{c\sqrt{1-m^{2}}}(\cos
ck_{z}-m)J_{z},  \label{cos}
\end{equation}%
where the Brillouin zone is taken as $-\pi /c\leq k_{z}\leq \pi /c$ with $c$
being the lattice constant. A pair of multi-band touching points exist at $%
ck_{z}=\pm \arccos m$ for $|m|<1$.

In the vicinity of these points, the Hamiltonian is expanded as%
\begin{equation}
H=\hbar vk_{x}J_{x}+\hbar vk_{y}J_{y}\pm \hbar v(k_{z}\pm \frac{\arccos m}{c}%
)J_{z}.
\end{equation}%
The $k_{z}$-dependent total Chern number now reads%
\begin{equation}
C\left( k_{z}\right) =\frac{N}{2}\text{sgn}\left( \cos ck_{z}-m\right) .
\end{equation}%
Integrating this over $k_{z}$, we obtain 
\begin{equation}
C=\int_{-\pi /c}^{\pi /c}C\left( k_{z}\right) dk_{z}=\frac{1}{2}Nb_{z},
\end{equation}%
where $b_{z}=\left( \arccos m-\pi \right) /c$ is the distance between the
two multi-band touching points. Note that $C$ is no longer a topological
charge and not quantized.

\begin{figure}[t]
\centerline{
\includegraphics[width=0.5\textwidth]{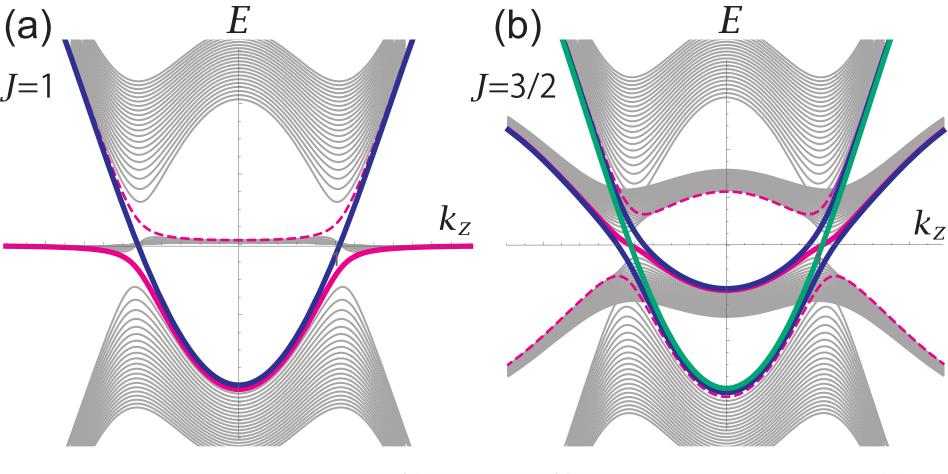}}
\caption{ Landau Levels for a pair of multi-band touching fermion
multiplets. (a) 3-band touching semimetal with $J=1$, where $N_{\text{ABJ}%
}=2 $. (b) 4-band touching semimetal with $J=3/2$, where $N_{\text{ABJ}}=4$.
See the caption of Fig.1.}
\label{FigCos}
\end{figure}

\textbf{Anomalous Hall effect:} The Thouless-Kohmoto-Nightingale-Nijs
formula is valid also in the present 3D case. Indeed, we first apply it in
the 2D space indexed by $k_{z}$, and then we integrate it over $k_{z}$ to
obtain the anomalous Hall effect as%
\begin{equation}
\mathbf{j}=N\frac{e^{2}}{2\pi ^{2}\hbar ^{2}}\mathbf{b}\times \mathbf{E}%
,\quad |\mathbf{b}|=\frac{\arccos m-\pi }{c},  \label{AnomaHC}
\end{equation}%
where the rotational invariance has been restored. The anomalous Hall
conductance becomes $N$ times larger than that of the Weyl semimetal.

\textbf{The Adler-Bell-Jackiew anomaly:} The analysis of the LLs is
important to reveal the ABJ anomaly in the system. In crystals multi-bands
touching points emerge in pairs. Making the minimum substituting into the
Hamiltonian (\ref{cos}) we obtain the LLs for such a pair and show them in
Fig.\ref{FigCos}(a). When we apply electric field along the $z$ direction
additionally, charge pumping occurs via the mid-gap LLs as in the Weyl
semimetal\cite{ABJ,Hosur}, which is a manifestation of the ABJ anomaly.
Namely, the chiral charge, which is defined by the difference of the charge
between the two multi-band touching points, is not conserved,%
\begin{equation}
\frac{\partial \rho _{\chi }}{\partial t}=\chi N_{\text{ABJ}}\frac{e^{3}}{%
h^{2}}\mathbf{E}\cdot \mathbf{B},
\end{equation}%
where $\chi =\pm 1$ denotes the chirality and $\rho _{\chi }$ is the charge
of each multi-band touching point, while $N_{\text{ABJ}}$ is the number of
channels. The mid-gap LLs which contribute to the ABJ anomaly are those
crossing the Fermi energy. The number of such LLs is $N_{\text{ABJ}}$. Here, 
$N_{\text{ABJ}}=2$ for $J=1$. It will be observed experimentally as an
enhancement of the negative magnetoresistance. This is consistent with the
result that the anomalous Hall conductance becomes $N$ times larger than
that of the Weyl semimetal with $N=2$ for $J=1$: See (\ref{AnomaHC}).
Namely, $N_{\text{ABJ}}=N$.

\textbf{Pair annihilation induced by photo-irradiation:} Creation of Weyl
semimetals by photo-irradiation has been discussed intensively \cite%
{PWang,PChan, Ebihara,PYan, PChan2}. We generalize the analysis to the
multi-band touching semimetals. We consider a beam of circularly polarized
light irradiated onto the multi-band touching semimetals. We take the
electromagnetic potential as $\mathbf{A}(t)=(A\cos \omega t,A\sin \left(
\omega t+\phi \right) ,0)$, where $\omega $ is the frequency of light with $%
\omega >0$ for the right circulation and $\omega <0$ for the left
circulation. The choice $\phi =0$ and $\phi =\pi $ corresponds to the
right-handed and left-handed circularly polarized light, respectively.

\begin{figure}[t]
\centerline{\includegraphics[width=0.5\textwidth]{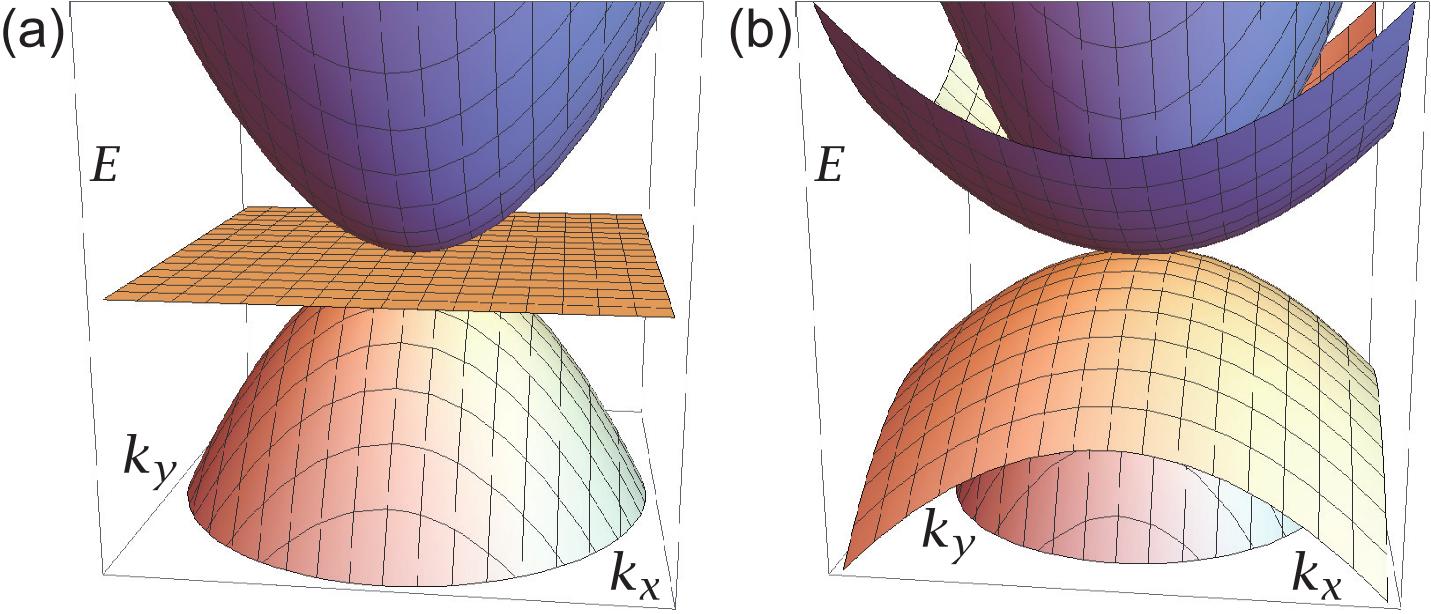}}
\caption{ Band structure of double $J$-fermions. The horizontal plane is
spanned by the $k_{x}$ and $k_{y}$ axes. The vertical axis is the energy $E$
as a function of $k_{x}$ and $k_{y}$, where we have set $k_{z}=0$.
Multi-band touching occurs quadratically for $N=2$. (a) $J=1$ and (b) $J=3/2$%
. We have set $\hbar v=c=1$.}
\label{FigDouble}
\end{figure}

We discuss the effect of photo-irradiation. The effective Hamiltonian is
given by\cite{Oka09L,Inoue,Kitagawa01B,Lindner,Dora,EzawaPhoto,Gold}%
\begin{equation}
\Delta H_{\text{eff}}=\frac{1}{\hbar \omega }\sum_{n\geq 1}\frac{\left[
H_{-n},H_{+n}\right] }{n},
\end{equation}%
with%
\begin{equation}
H_{+n}=\frac{1}{T}\int_{0}^{T}He^{in\omega t}dt.
\end{equation}%
By using%
\begin{eqnarray}
H_{+1} &=&\hbar veA\left( \frac{1+e^{-i\phi }}{2}\right) J_{+},\qquad \\
H_{-1} &=&\hbar veA\left( \frac{1+e^{i\phi }}{2}\right) J_{-},
\end{eqnarray}%
and $H_{\pm n}=0$ for $n\geq 2$, the effective Hamiltonian is calculated as%
\begin{equation}
\Delta H_{\text{eff}}=\frac{1}{\hbar \omega }\left[ H_{-1},H_{+1}\right] =-%
\frac{2\left( \hbar veA\right) ^{2}}{\hbar \omega }J_{z}\cos \phi ,
\label{Photo}
\end{equation}%
which only shifts the crossing point in the $z$ direction $k_{z}\mapsto
k_{z}-2e^{2}A^{2}/\left( \hbar ^{2}\omega \right) $ for the Hamiltonian (\ref%
{kJ}). By adding the term (\ref{Photo}) to the Hamiltonian (\ref{cos}), we
find the parameter $m$ is renormalized to be $m+\frac{2e^{2}A^{2}}{\hbar
\omega }$. We investigate the energy spectrum by changing $A$. When $%
\left\vert m+\frac{2e^{2}A^{2}}{\hbar \omega }\right\vert =1$, a pair
annihilation occurs between two multi-band touching points, as illustrated
in Fig.\ref{FigPhoto}(a) and (b) for the cases of $J=1$ and $3/2$. A
detached perfect flat band appears for integer $J$ after the pair
annihilation, as in Fig.\ref{FigPhoto}(c).

\begin{figure}[t]
\centerline{\includegraphics[width=0.5\textwidth]{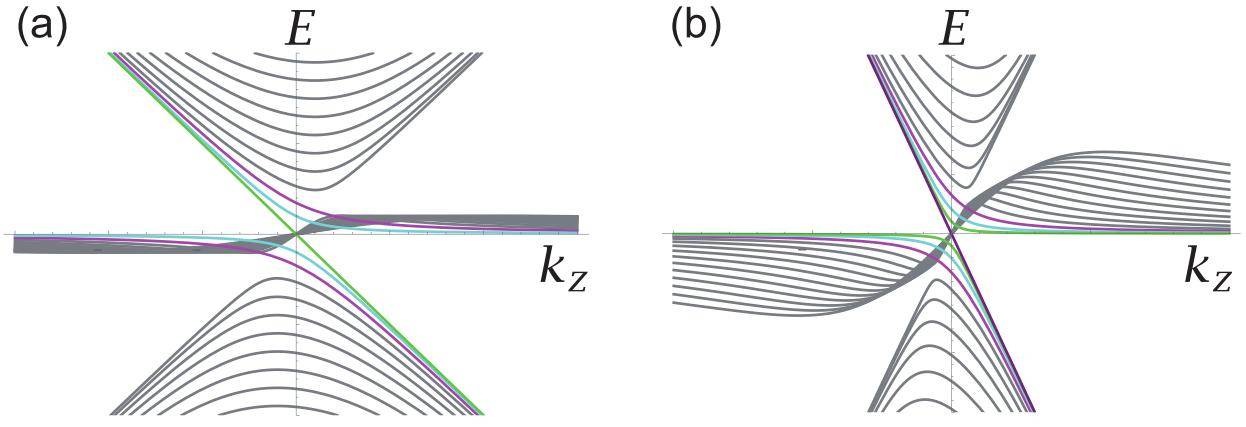}}
\caption{ Landau Levels for a multiple $J$-fermions with $J=1$. (a) double $J
$-fermions, (b) triple $J$-fermions, See the caption of Fig.1.}
\label{FigJLL}
\end{figure}

\section{Multiple $J$-fermions}

\textbf{Hamiltonian:} Multiple-Weyl semimetals are described by\cite%
{CFang,BJYang,Xli,Huang2}%
\begin{equation}
H=\left( 
\begin{array}{cc}
\hbar vk_{z} & \frac{c}{2}k_{-}^{N} \\ 
\frac{c}{2}k_{+}^{N} & -\hbar vk_{z}%
\end{array}%
\right) =\frac{c}{2}k_{+}^{N}\sigma _{-}+\frac{c}{2}k_{-}^{N}\sigma
_{+}+\hbar vk_{z}\sigma _{z}  \label{MHPF}
\end{equation}%
with an integer $N$. It is known in crystals that only double-Weyl
semimetals with $N=2$ and triple-Weyl semimetals with $N=3$ are possible due
to the crystal symmetry restriction\cite{CFang,BJYang}.

We propose to generalize it for $J$-fermions as%
\begin{equation}
H=\frac{c}{2}k_{+}^{N}J_{-}+\frac{c}{2}k_{-}^{N}J_{+}+\hbar vk_{z}J_{z}.
\end{equation}%
We call them double $J$-fermions for $N=2$ and triple $J$-fermions for $N=3$%
, as in the case of double Weyl semimetals and triple Weyl semimetals.

Using the polar coordinate, the Hamiltonian is rewritten as%
\begin{equation}
H=ck^{N}\left( J_{x}\cos N\phi +J_{y}\sin N\phi \right) +\hbar vk_{z}J_{z},
\end{equation}%
the Hamiltonian is diagonalized as 
\begin{equation}
UHU^{-1}=\sqrt{c^{2}\left( k_{x}^{2}+k_{y}^{2}\right) ^{N}+\left( \hbar
vk_{z}\right) ^{2}}J_{z},
\end{equation}%
where 
\begin{equation}
U=\exp [i\theta J_{x}]\exp [i\left( N\phi +\frac{\pi }{2}\right) J_{z}],
\end{equation}%
with $\tan \theta =\frac{ck^{N}}{\hbar vk_{z}}$. The energy dispersion is
obtained as%
\begin{equation}
E=j\sqrt{c^{2}\left( k_{x}^{2}+k_{y}^{2}\right) ^{N}+\left( \hbar
vk_{z}\right) ^{2}}.
\end{equation}%
We show the band structures of double $J$-fermion with $J=1$ and $J=3/2$ in
Fig.\ref{FigDouble}. The band touching is not linear but quadratic.

\textbf{Monopole charges: } It is known that the monopole charges for double
and triple Weyl semimetals are given by $\pm 2$ and $\pm 3$ respectively. We
generalize the results to the case of general $J$-fermions. Using the
relations%
\begin{equation}
\frac{\partial U}{\partial \theta }\frac{\partial U^{-1}}{\partial \phi }-%
\frac{\partial U}{\partial \phi }\frac{\partial U^{-1}}{\partial \theta }%
=NJ_{z}\sin \theta -NJ_{y}\cos \theta ,
\end{equation}%
the Berry curvature is explicitly calculated for each band as%
\begin{equation}
\mathbf{\Omega }_{j}=Nj\frac{\mathbf{k}}{k^{3}}.
\end{equation}%
Since $\frac{1}{2\pi }\int d^{3}k\,\nabla \cdot \mathbf{\Omega }_{j}=2Nj$,
the Berry curvature of the band indexed by $j$ describes a monopole with the
monopole charge $2Nj$ in the momentum space.

\begin{figure}[t]
\centerline{\includegraphics[width=0.5\textwidth]{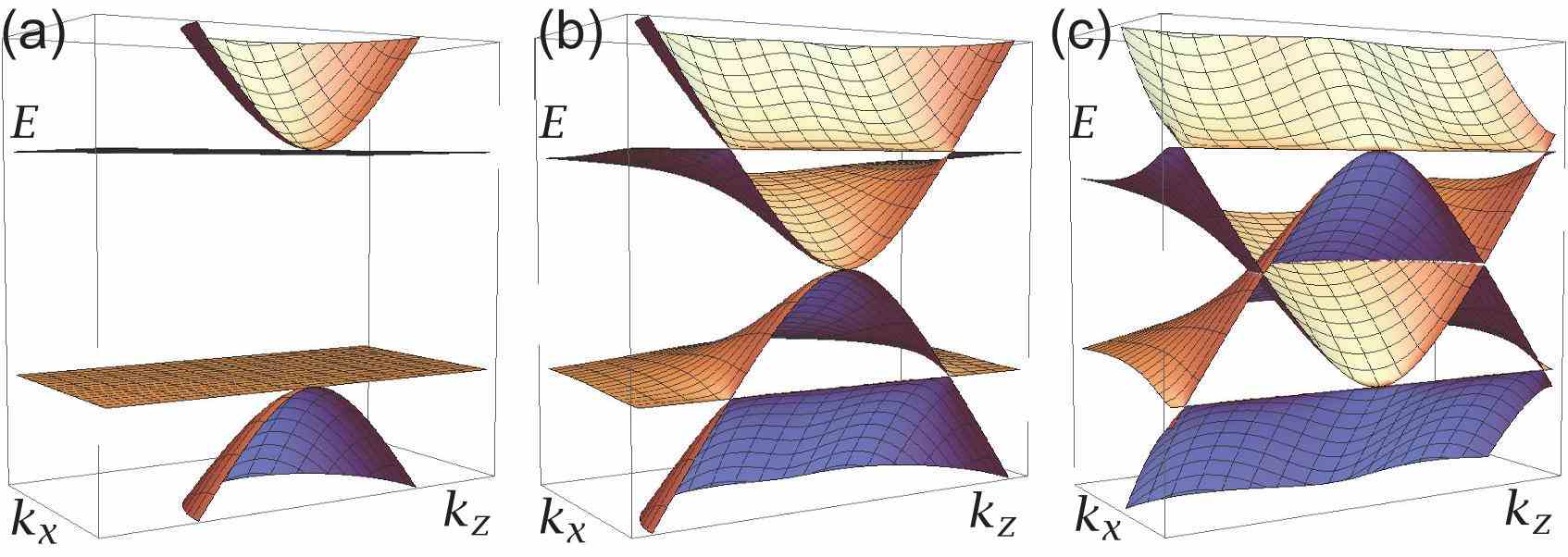}}
\caption{ Band structure of Dirac-like $J$-fermions with $J=1$ under
photo-irradiation. The horizontal plane is spanned by the $k_{z}$ and $k_{x}$
axes. The vertical axis is the energy $E$ as a function of $k_{z}$ and $k_{x}
$, where we have set $k_{y}=0$. (a) $h=0$, (b) $h=m$ and (c) $h=2m$. (b)
represents a critical point of a Lifshitz transition. }
\label{FigJ3H}
\end{figure}

\textbf{Landau levels:} The Hamiltonian under magnetic field is given by%
\begin{equation}
\hat{H}=\frac{\hbar \omega _{\text{c}}}{2}\left( \hat{a}^{\dagger }\right)
^{^{N}}J_{-}+\frac{\hbar \omega _{\text{c}}}{2}\hat{a}^{N}J_{+}+\hbar
vk_{z}J_{z}.
\end{equation}%
The bulk spectrum is obtained by setting 
\begin{equation}
\psi =\left( u_{n}\left\vert n\right\rangle ,u_{n+N}\left\vert
n+N\right\rangle ,u_{n+2N}\left\vert n+2N\right\rangle \right) ^{t}
\end{equation}%
with $J=1$ and 
\begin{equation}
\psi =\left( u_{n}\left\vert n\right\rangle ,\cdots ,u_{n+JN}\left\vert
n+JN\right\rangle \right) ^{t}
\end{equation}%
with general $J$ for $n\geq 0$ and solving the eigenvalue problem $\hat{H}%
\psi =E\psi $. We show the LLs in Fig.\ref{FigJLL}. The number of the
mid-gap LLs is $N_{\text{mid}}=2JN(2J+1)/2$.

\begin{figure*}[t]
\centerline{\includegraphics[width=\textwidth]{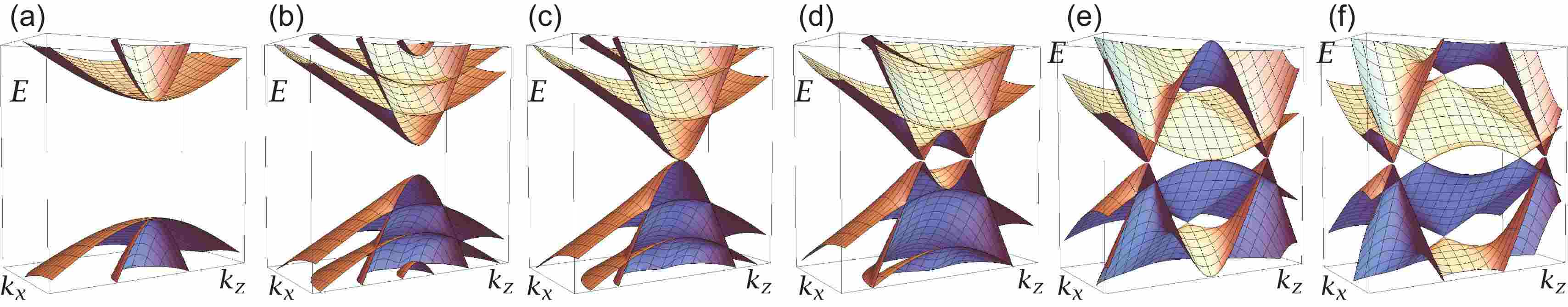}}
\caption{ Band structure of Dirac-like $J$-fermions with $J=3/2$ under
photo-irradiation. The horizontal plane is spanned by the $k_{z}$ and $k_{x}$
axes. The vertical axis is the energy $E$ as a function of $k_{z}$ and $k_{x}
$, where we have set $k_{y}=0$. (a) $h=0$, (b) $h=m/2$, (c) $h=2m/3$, (d) $%
h=m$, (e) $h=2m$ and (f) $h=5m/2$. (c) and (e) represent critical points of
Lifshitz transitions. }
\label{FigJ4H}
\end{figure*}

\textbf{Photo-irradiation:} The effective Hamiltonian with photo-irradiation
is given by 
\begin{equation}
\Delta H_{\text{eff}}=-2\frac{\left( c/2\right) ^{2}}{\hbar \omega }%
\sum_{n=0}^{N}\frac{\left( _{N}C_{n}\right) ^{2}}{n}\left( eA\right)
^{2n}\left( k_{x}^{2}+k_{y}^{2}\right) ^{N-n}J_{z},
\end{equation}%
where we have used%
\begin{eqnarray}
H_{+n} &=&_{N}C_{n}\frac{c}{2}\left( eA\right) ^{n}k_{-}^{N-n}J_{+},\qquad \\
H_{-b} &=&_{N}C_{n}\frac{c}{2}\left( eA\right) ^{n}k_{+}^{N-n}J_{-}.
\end{eqnarray}%
The band degeneracy is not resolved but only the dispersion shifts and
changes.

\section{Dirac-like $J$-fermions}

\textbf{Hamiltonian:} In the chiral representation, the Dirac fermion is
described by the Hamiltonian%
\begin{equation}
H=\hbar v\mathbf{k}\cdot \mathbf{\sigma }\tau _{z}+m\tau _{x}.
\end{equation}%
We propose to generalize it to $J$-fermions as%
\begin{equation}
H=\hbar v\mathbf{k}\cdot \mathbf{J}\tau _{z}+m\tau _{x},
\end{equation}%
which yields the energy spectrum%
\begin{equation}
E=j\sqrt{\left( \hbar vk\right) ^{2}+m^{2}}.
\end{equation}%
When $m=0$, this Hamiltonian is split into two independent Hamiltonians
describing the $J$-fermions whose monopole charges are opposite.

\textbf{Photo-irradiation:} Especially, the perfect flat bands emerge at $%
E=\pm m$ for the even $J$. The effective Hamiltonian by photo-irradiation is
calculated as%
\begin{equation}
\Delta H_{\text{eff}}=\frac{1}{\hbar \omega }\left[ H_{-1},H_{+1}\right] =-%
\frac{2\left( \hbar veA\right) ^{2}}{\hbar \omega }J_{z}\cos \phi ,
\end{equation}%
where we have used%
\begin{eqnarray}
H_{+1} &=&\hbar veA\left( \frac{1+e^{-i\phi }}{2}\right) J_{+}\tau _{z}, \\
H_{-1} &=&\hbar veA\left( \frac{1+e^{i\phi }}{2}\right) J_{-}\tau _{z},
\end{eqnarray}%
and $H_{\pm n}=0$ for $n\geq 2$. The Hamiltonian is summarized as%
\begin{equation}
H=\hbar v\mathbf{k}\cdot \mathbf{J}\tau _{z}+m\tau _{x}+hJ_{z}
\end{equation}%
with $h=-\frac{2\left( \hbar veA\right) ^{2}}{\hbar \omega }\cos \phi $.

It is hard to obtain the energy dispersion for general $\mathbf{J}$.
However, we can diagonalize the Hamiltonian along the $k_{z}$ axis, for
which $k_{x}=k_{y}=0$. Wit the use of the unitary transformation%
\begin{equation}
U=\exp [i\theta \tau _{y}],\quad \tan \theta =\frac{m}{\hbar vk_{z}},
\end{equation}%
we obtain 
\begin{equation}
U^{-1}\tau _{z}U=\tau _{x}\sin \theta +\tau _{z}\cos \theta ,
\end{equation}%
and hence 
\begin{equation}
UH\left( 0,0,k_{z}\right) U^{-1}=\sqrt{\left( \hbar vk_{z}J_{z}\right)
^{2}+m^{2}}\tau _{z}+hJ_{z}.
\end{equation}%
The energy along the $k_{z}$ axis is found to be%
\begin{equation}
E=t\sqrt{\left( \hbar vk_{z}J_{z}\right) ^{2}+m^{2}}+hj,
\end{equation}%
where $t=\pm 1$ and $j=-J,-J+1,\cdots ,J-1,J$. There are several Lifshitz
transitions at 
\begin{equation}
h=\frac{t}{j}m,
\end{equation}%
where the band gap closes at $\mathbf{k}=\mathbf{0}$. For example, there are
transitions at $h=\pm m$ for $J=1$ and $h=\pm 2m/3$ and$h=\pm 2m$ for $J=3/2$%
. We show the numerically obtained band structure for various $h$ in Figs.%
\ref{FigJ3H} and \ref{FigJ4H}.

\section{Loop-nodal $J$-fermions}

\begin{figure}[t]
\centerline{\includegraphics[width=0.5\textwidth]{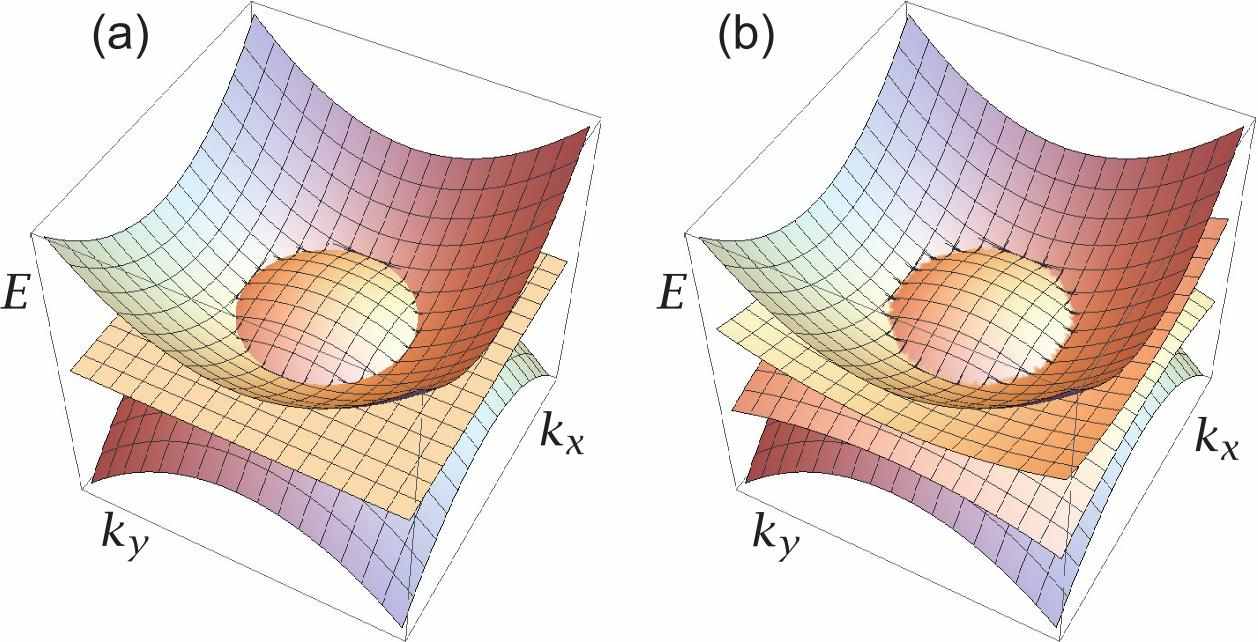}}
\caption{Band structure of loop-nodal $J$-fermions with (a) $J=1$ and (b) $%
J=3/2$. The horizontal plane is spanned by the $k_{x}$ and $k_{y}$ axes. The
vertical axis is the energy $E$ as a function of $k_{x}$ and $k_{y}$, where
we have set $k_{z}=0$. A loop node exists along the $k_{x}$-$k_{y}$ plane
for $m/a>0$. We have set $\hbar v=m=a=1$.}
\label{FigLoop}
\end{figure}

\textbf{Hamiltonian:} In the 3D space, it is known that another type of
semimetal called loop-nodal semimetals is possible. Loop-nodal semimetals
are described by the Hamiltonian\cite%
{Burkov,Carter,Phillips,ChenLu,Zeng,Chiu,Mullen,Weng,Yu,Kim,Bian,Xie,Hyper,Rhim,ChenXie,Fang,BianChang,Chan,RhimKim,Yan}%
\begin{equation}
H=\left[ a\left( k_{x}^{2}+k_{y}^{2}\right) -m\right] \sigma
_{x}+ck_{z}\sigma _{z}.
\end{equation}%
We propose to generalize it to $J$-fermions as%
\begin{equation}
H=\left[ a\left( k_{x}^{2}+k_{y}^{2}\right) -m\right] J_{x}+ck_{z}J_{z}.
\end{equation}%
Using the relation%
\begin{equation}
U^{-1}J_{z}U=J_{x}\sin \theta +J_{z}\cos \theta ,\quad \tan \theta =\frac{%
a\left( k_{x}^{2}+k_{y}^{2}\right) -m}{ck_{z}}
\end{equation}%
with $U=\exp [i\theta J_{y}]$, the Hamiltonian is diagonalized as $%
UHU^{-1}=EJ_{z}$ with the energy spectrum%
\begin{equation}
E=j\sqrt{\left[ a\left( k_{x}^{2}+k_{y}^{2}\right) -m\right] ^{2}+\left(
ck_{z}\right) ^{2}}.
\end{equation}%
We show the band structure of loop-nodal $J$-fermions in Fig.\ref{FigLoop}. 
The gap closes at the nodal loop 
\begin{equation}
k_{x}^{2}+k_{y}^{2}=\frac{m}{a}
\end{equation}%
for $m/a>0$. This loop node is protected by the mirror symmetry%
\begin{equation}
MH\left( k_{x},k_{y},k_{z}\right) M^{-1}=H\left( k_{x},k_{y},-k_{z}\right) 
\end{equation}%
with the mirror operator $M=iJ_{x}$.

\begin{figure}[t]
\centerline{\includegraphics[width=0.5\textwidth]{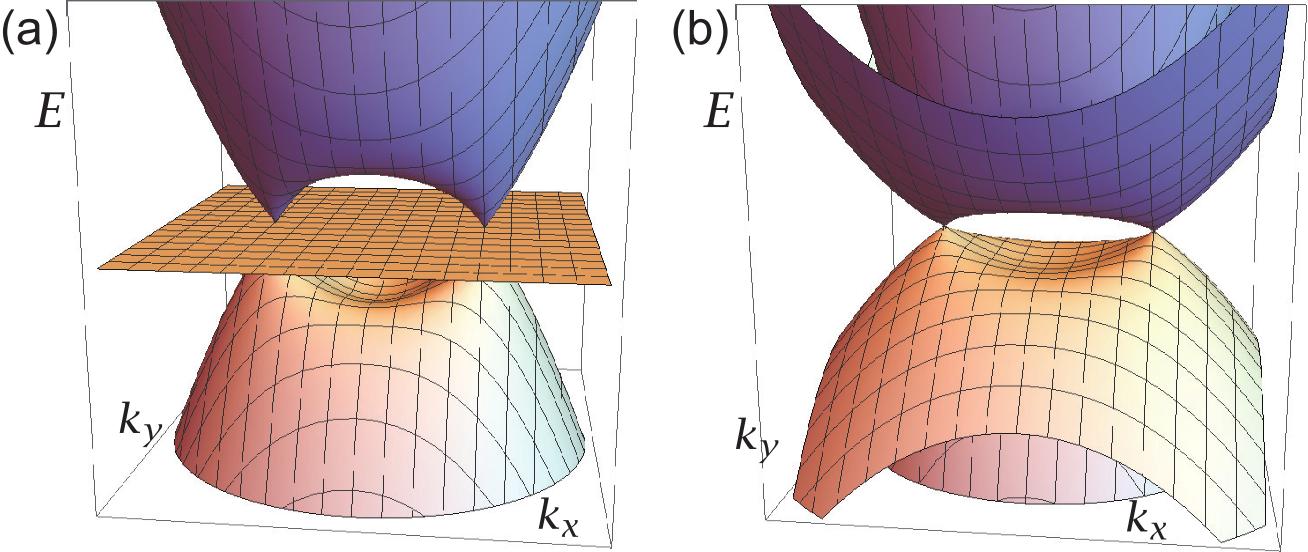}}
\caption{ Band structure of loop-nodal $J$-fermions with photo-irradiation.
The horizontal plane is spanned by the $k_{x}$ and $k_{y}$ axes. The
vertical axis is the energy $E$ as a function of $k_{x}$ and $k_{y}$, where
we have set $k_{z}=0$. Two $J$-fermions emerge by photo-irradiation along
the $x$ direction. The gap opens except for two points. (a) $J=1$ and (b) $%
J=3/2$. We have set $\hbar v=m=ac\left( eA\right) ^{2}\cos \protect\phi =1$.}
\label{FigPhotoLoop}
\end{figure}

\textbf{Photo-irradiation:} It is shown that the Weyl semimetal emerges when
we apply photo-irradiation along the $x$ direction to the loop-nodal
semimetals\cite{PYan}. We generalize it to loop-nodal $J$-fermions. For the
photo-irradiation along the $x$ direction with $\mathbf{A}(t)=(0,A\cos
\omega t,A\sin \left( \omega t+\phi \right) )$, the first order effective
Hamiltonian for (\ref{C3H}) is obtained as%
\begin{equation}
H^{\left( 1\right) }=-ac\left( eA\right) ^{2}k_{y}J_{y}\cos \phi .
\end{equation}%
The energy is modified as%
\begin{equation}
E=j\sqrt{\left[ a\left( k_{x}^{2}+k_{y}^{2}\right) -m\right] ^{2}+\left(
ck_{z}\right) ^{2}+\left( ac\left( eA\right) ^{2}k_{y}\cos \phi \right) ^{2}}%
.
\end{equation}%
The gap opens except for the two points 
\begin{equation}
k_{x}^{\pm }=\pm \sqrt{\frac{m}{a}},\quad k_{y}=0,\quad k_{z}=0.
\end{equation}%
In the vicinity of these points, the Hamiltonian is expanded as%
\begin{equation}
H=\mp 2\sqrt{am}\left( k_{x}\mp \sqrt{\frac{m}{a}}\right) J_{x}-ac\left(
eA\right) ^{2}k_{y}J_{y}\cos \phi +ck_{z}J_{z},
\end{equation}%
which shows that anisotropic $J$-fermions are realized by photo-irradiation.
We show the band structure in Fig.\ref{FigPhotoLoop}.

\section{C$_{3}$-protected fermions}

\textbf{Hamiltonian:} It has been proposed\cite{Chang,HWeng} that a
three-fold degeneracy is protected by the C$_{\mathbf{3}}$ symmetry and the
mirror symmetry $M_{y}$ of the Hamiltonian, 
\begin{equation}
H=\left( 
\begin{array}{ccc}
\hbar vk_{z} & \lambda _{1}k_{+}^{2} & \lambda _{2}k_{+} \\ 
\lambda _{1}k_{-}^{2} & \hbar vk_{z} & \lambda _{2}k_{-} \\ 
\lambda _{2}k_{-} & \lambda _{2}k_{+} & -\hbar vk_{z}%
\end{array}%
\right)  \label{C3H}
\end{equation}%
with $k_{\pm }=k_{x}\pm ik_{y}$. The energy spectrum is given by%
\begin{eqnarray}
E_{0} &=&\hbar vk_{z}-\lambda _{1}k^{2},  \notag \\
E_{\pm } &=&\frac{\lambda _{1}}{2}k^{2}\pm \sqrt{\left( 2\hbar
vk_{z}+\lambda _{1}k^{2}\right) ^{2}+8\lambda _{2}^{2}k^{2}}.
\end{eqnarray}%
A 3-band touching point exists at $\mathbf{k}=0$. In addition, there is a
line degeneracy along the line $k_{x}=k_{y}=0$. Thus the Chern number is
ill-defined. On the other hand, the Berry phase is nonzero for the $E_{0}$
band, 
\begin{equation}
\Gamma _{\text{B}}=-i\int d\theta \left\langle \psi _{0}\right\vert \frac{%
\partial }{\partial \theta }\left\vert \psi _{0}\right\rangle =2\pi ,
\end{equation}%
while it is zero for the $E_{\pm }$ bands.

\begin{figure}[t]
\centerline{\includegraphics[width=0.5\textwidth]{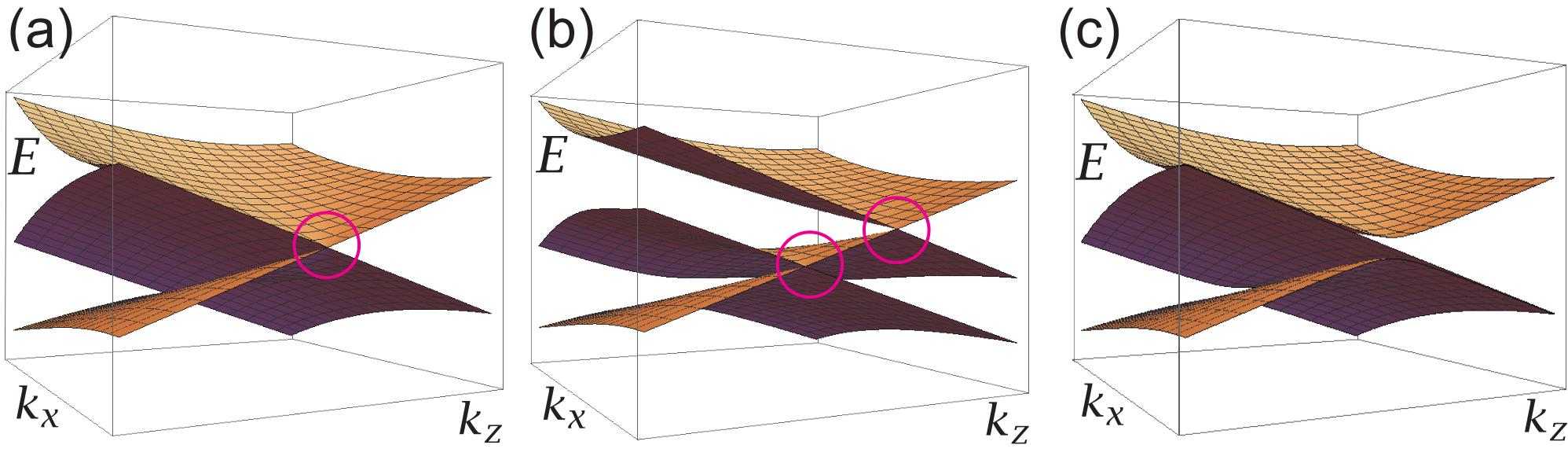}}
\caption{Breaking of C$_{3}$-protected fermion by photo-irradiation. The
vertical axis is the energy $E$ as a function of $k_{z}$ and $k_{x}$ by
setting $k_{y}=0$.\ (a) C$_{3}$-protected fermion without photo-irradiation.
The 3-fold degenerate point is marked by a magenta circle. (b) That with
photo-irradiation along the $z$ direction and (c) along the $x$ direction.
The 3-fold degenerate point is split into two Weyl points marked by magenta
circles. }
\label{FigCPhoto}
\end{figure}

\textbf{Photo-irradiation:} We study photo-irradiation effects. We show that
the three-band touching is resolved since photo-irradiation breaks the
mirror symmetry.

For the photo-irradiation along the $z$ direction with $\mathbf{A}(t)=(A\cos
\omega t,A\sin \left( \omega t+\phi \right) ,0)$, the first order effective
Hamiltonian for (\ref{C3H}) is obtained as%
\begin{equation}
H^{\left( 1\right) }=\cos \phi \left( 
\begin{array}{ccc}
-F & 0 & -G \\ 
0 & F & G^{\ast } \\ 
-G^{\ast } & G & 0%
\end{array}%
\right) 
\end{equation}%
with 
\begin{eqnarray}
F &=&\frac{1}{\hbar \omega }\left( eA\right) ^{2}\left( \lambda
_{2}^{2}+4\lambda _{1}^{2}q^{2}\right) ,  \notag \\
G &=&\frac{2}{\hbar \omega }\left( eA\right) ^{2}\lambda _{1}\lambda
_{2}q_{+}.
\end{eqnarray}%
On the other hand, for the second order term we obtain%
\begin{equation}
H^{(2)}=\text{diag.}\left( F^{\prime },-F^{\prime },0\right) \cos \phi 
\end{equation}%
with 
\begin{equation}
F^{\prime }=\frac{1}{\hbar \omega }\left( eA\right) ^{2}\lambda _{1}^{2},
\end{equation}%
and $H_{\pm n}=0$ for $n\geq 3$. The terms proportional to $F$ breaks the
3-fold degeneracy along the C$_{3}$ invariant line $k_{x}=k_{y}=0$, while
the terms proportional to $G$ renormalize the velocity. As a result, the
3-band touching point is split into two Weyl points,%
\begin{equation}
H=\varepsilon _{\pm }+\lambda _{2}\sigma _{x}\pm \lambda _{2}\sigma
_{y}+v\sigma _{z},
\end{equation}%
by the photo-irradiation as shown in Fig.\ref{FigCPhoto}(b), which is highly
contrasted to the case of the $J$-fermions.

For the photo-irradiation along the $x$ direction with 
\begin{equation}
\mathbf{A}(t)=(0,A\cos \omega t,A\sin \left( \omega t+\phi \right) ),
\end{equation}
the first order effective Hamiltonian for (\ref{C3H}) is obtained as%
\begin{equation}
H^{\left( 1\right) }=F^{\prime }\left( 
\begin{array}{ccc}
0 & 0 & -1 \\ 
0 & 0 & 1 \\ 
-1 & 1 & 0%
\end{array}%
\right) ,
\end{equation}%
with 
\begin{equation}
F^{\prime \prime }=\cos \phi \frac{4}{\hbar \omega }\hbar v\lambda _{2},
\end{equation}
and $H_{\pm n}=0$ for $n\geq 2$. The 3-band touching is resolved as shown in
Fig.\ref{FigCPhoto}(c).

\section{Conclusion}

We have studied two typical types of multi-band touching fermions. One is
described by $J$-fermions described by the pseudospin operators in $J$%
-dimensional representation, where the chiral anomaly is induced by the
presence of monopoles. As experimental evidences the enhancement of the
anomalous Hall effect and the negative magnetoresistance are expected.
Photo-irradiation does not resolve the degeneracy but shifts the position of
the multi-band touching points, which results in the pair annihilation of
them. The other is a 3-band touching fermion protected by the C$_{3}$
symmetry. Photo-irradiation breaks the 3-band touching point into two Weyl
points. We have also proposed generalizations of Dirac semimetals,
multiple-Weyl semimetals and loop-nodal semimetals to those composed of $J$%
-fermions.

The author is very much grateful to N. Nagaosa and H. Ding for many helpful
discussions on the subject. He thanks the support by the Grants-in-Aid for
Scientific Research from MEXT KAKENHI (Grant Nos.JP25400317 and JP15H05854).
This work was also supported by CREST, JST.

\end{document}